\documentclass[12pt]{article}
\usepackage[pctex32]{graphics}
\begin{document}
\vspace{2cm}
\begin{center}
{\bf  \Large   Anisotropic Cosmological Models with Energy Density Dependent Bulk Viscosity}
\vspace{1cm}

                      Wung-Hong Huang\\
                       Department of Physics\\
                       National Cheng Kung University\\
                       Tainan,70101,Taiwan\\

\end{center}
\vspace{2cm}
    An analysis is presented of the Bianchi type I cosmological models with a bulk viscosity when the universe is filled with the stiff fluid $p = \epsilon$ while the viscosity is a power function of the energy density, such as $\eta = \alpha |\epsilon|^n$.  Although the exact solutions are obtainable only when the $2n$ is an integer, the characteristics of evolution can be clarified for the models with arbitrary value of $n$. It is shown that, except for the $n = 0$ model that has solutions with infinite energy density at initial state, the anisotropic solutions that evolve to positive Hubble functions in the later stage will begin with Kasner-type curvature singularity and zero energy density at finite past for the $n> 1$ models, and with finite Hubble functions and finite negative energy density at infinite past for the $n < 1$ models.  In the course of evolution, matters are created and the anisotropies of the universe are smoothed out. At the final stage, cosmologies are driven to infinite expansion state, de Sitter space-time, or Friedman universe asymptotically. However, the de Sitter space-time is the only attractor state for the $n <1/2 $ models. The solutions that are free of cosmological singularity for any finite proper time are singled out. The extension to the higher-dimensional models is also discussed.

\vspace{3cm}
\begin{flushleft}
E-mail:  whhwung@mail.ncku.edu.tw\\
Publised in: J. Math. Phys. 31 (1990) 1456-1462
\end{flushleft}


\newpage
\section{Introduction}
   The investigation of relativistic cosmological models usually has the energy momentum tensor of matter as that due to a perfect fluid. To consider more  realistic models one must take into account the viscosity mechanisms, which
have already attracted the attention of many investigators. Misner [1] suggested that strong dissipative due to the neutrino viscosity may considerably reduce the anisotropy of the blackbody radiation. Viscosity mechanism in the cosmology
can explain the anomalously high entropy per baryon in the present universe [2,3].  Bulk viscosity associated with the grand-unified-theory phase transition [4] may lead to an inflationary scenario [5-7].
   
   An exactly soluble isotropic cosmological model of the zero curvature Friedman model in the presence of bulk viscosity has been examined by Murphy [8].  The solutions that he found exhibit an interesting feature that the big bang type
singularity appears in infinite past.  Exact solutions of the isotropic homogeneous cosmology for the open, closed and flat universe have been found by Santos et al .[9], when the bulk viscosity is the power function of energy density.  However,
in some cases, the big bang singularity occurs at finite past.  It is thus shown that Murphy's conclusion that the introduction of bulk viscosity can avoid the initial singularity at finite past is not, in general, valid. (The extensive collections of
exact isotropic solutions are those found in Ref. 10.)

    Belinskii and Khalatnikov [11] analyzed the Bianchi type I cosmological models under the influence of viscosity. They then found the remarkable property that near the initial singularity the gravitional field creates matters. Using a certain simplifying assumption, Banerjee and Santos [12-13] obtained some exact solutions for the homogeneous anisotropic model. Recently Banerjee et al.[14] obtained some Bianchi type I solutions for the case of stiff matter by using the assumption that shear viscosity is the power function of the energy density. However, the bulk viscosity coefficients adopted in their model are zero or constant. 

    In this paper, without introducing the shear viscosity, we shall examine the Bianchi type I cosmological models with bulk viscosity  ($\eta$) that is a power function of energy density ($\epsilon$), i.e., $\eta = \alpha |\epsilon|^n$, when the universe is filled with the stiff matter $p=\epsilon$. We are interested in the cosmological solutions that will eventually go to the states of positive Hubble functions. The exact solutions are obtained when $n$ is an integer.  Furthermore, through some analyses, we can know how evolutions of the models with arbitrary value of  $n$ will be.  We prove that the isotropic de Sitter space-time is a stable attractor state as $t \rightarrow \infty$ if $n < 1/2$. It is thus in accord with the "cosmic no hair" theorem [15-17] even though the strong energy condition [18] is violated [19]. (The weak energy condition question in the anisotropic viscous models has been discussed by Barrow [20].)  

   We also find that, for the models of $n < l $ there are solutions (in fact, all solutions in $0 < n <1/2$ models) that can avoid the cosmological singularity at any finite proper time. Our models show that the anisotropies of the universe are smoothed out and matters are created by the gravitational field in the course of the evolution, in agreement with the results found by others [11,14]. 

   The models of $n = 0, 1$ have been discussed in our previous paper [21]. However, the method invented there cannot be used to solve the models with other $n$, and the analyses of the $n = 0$ model were incomplete. Barrow [22] had also given further discussion of bulk viscous models in theories possessing quadratic curvature. It is worth mentioning that the $n = 3/2$ model can be used to describe the quantum production of infinitely thin Witten strings [23] on super-horizon scales in the very early universe (see the arguments in the paper of Turok [24]), which has been analyzed by Barrow recently [10].

The outline of this paper is as follows. We first discuss the axially symmetric Bianchi type I model in which there are ony two cosmic scale functions. In Sec. II we reduce the Einstein's field equation to a pair of coupled differential equations that become an integratable equation when we define two suitable variables in Sec. III. The exact solutions for the models with integer In are then found. The analyses about the initial and final states of the models with any $n$ are given
in Sec. IV. To discuss the generic Bianchi type I model with multiple cosmic scale functions we then in Sec. V present a simple method that can also be used to analyze the higher-dimensional models. Section VI is devoted to conclusions.

\section{ Einstein's  Field Equation and Some Analyses}
   We first consider the axially symmetric Bianchi type I model with a metric in the form 

$$ds^2=g_{\mu\nu}dx^\mu dx^\nu = -dt + X(t)^2dx^2 + Y(t)^2 (dy^2+dz^2),  \eqno{(2.1)}$$
\\
where $X$ and $Y$ are functions of cosmic time $t$ alone. The field equations to be solved are

$$R_{\mu\nu}=((\epsilon -\bar p)/2) g_{\mu\nu}+(\epsilon +\bar p)u_\mu u_nu,   \eqno{(2.2)}$$
\\
where $\epsilon$ is the energy density, and $u_\mu$ is the four-velocity that satisfies
$$u_\mu u^\mu = -1 .  \eqno{(2.3)}$$

The total pressure $\bar p$ is defined by

$$\bar p = p- \eta u^\mu_{;\mu},     \eqno{(2.4)}$$
\\
where $p$ is the pressure coming from the perfect fluid and $\eta$ is the bulk viscosity. Choosing a comoving frame where $u_\mu = \delta ^0_\mu$ , the Einstein's field equations (2.2) lead to 

$${dH\over dt} + H W = (\epsilon -\bar p)/2,  \eqno{(2.5)}$$
$${dh\over dt} + h W = (\epsilon -\bar p)/2,  \eqno{(2.6)}$$
$$W^2 - (2H^2 + h^2) =2 \epsilon,  \eqno{(2.7)}$$
\\
where $H$ and $h$ are the Hubble functions defined by

$$H = {dY\over dt} Y^{-1},   \eqno{(2.8a)}$$
$$h = {dX\over dt} X^{-1},   \eqno{(2.8b)}$$
\\
and $W$ is the total expansion function
$$W = 2H + h.                                \eqno{(2.8c)}$$
We also have a relation

$$\epsilon-\bar p = (2-\gamma )\epsilon +\eta W,  \eqno{(2.9)}$$
\\
where $\gamma$ is defined by the equation of state

$$p=(\gamma - 1)\epsilon , ~~~ l < \gamma < 2.           \eqno{(2.10)}$$

In this paper we only consider the stiff matter, i.e., $\gamma = 2$, which is the possible relevance of the equation of state $p= \epsilon$ regarding the matter content of the early universe [25-26].  Using the assumption that the bulk viscosity is a power function of energy density, 

$$\eta = \alpha \epsilon^n   , n\geq 0.            \eqno{(2.11)}$$

Equations (2.5) and (2.6) then become

$${dH\over dt} + H W = {\alpha \over 2}[H(H+2h)]^n W,           \eqno{(2.12)}$$
$${dh\over dt} + h W = {\alpha \over 2}[H(H+2h)]^n W.           \eqno{(2.13)}$$

The work that remains is to analyze the above coupled differential equations -and find their exact solutions. We will express these solutions as the flows in the phase space $H \times h$ and thus find the dynamical evolutions of cosmology. We only consider the physical plane where the trajectories shall evolve to the positive Hubble functions in the latter stage.  The regions where $\epsilon < 0$  that violate the weak energy condition[18] are needed for the solutions of $n < 1$ models as discussed in next section. We then find there that the evolutions of the cosmology are confined in the regions of  $H>0$ and $H + 2h>0$ for the $n \geq1$ models, while these for the $n < 1$ models are $W\geq 0$.

Let us first determine the singular points, fixed points, and cosmic time in the phase space.

{\bf A. Singular points }

   Equation (2.7) tells us that the energy density becomes infinite only if $H$ and/or  $h$ are infinite. From Eqs. (2.12) and (2.13) we also know that $dH /dt$ and $dh/dt$ can be infinite only if $H$ and/or $h$ are infinite. As the Riemann scalar curvature can be written as 

$$R=2{dW\over dt} +W^2 +2H + h^2$$
\\
we see that $R$ can be infinitely large only if $H$ and/or $h$ are infinite. Therefore, {\it the singularity of diverge $R$ and $\epsilon$ could occur only if $H$ and /or $h$ is infinite.}

{\bf B. Fixed points}

   The fixed points are the solutions of Eqs. (2.12) and (2.13) once we let $dH/dt = dh /dt = 0$. They are

$$(1) W=0 => H=h=0~~~ or~~~ 2H=-h \not=0,  \eqno{(2.14a)}$$
$$(2) H=h =H_D \equiv [(3^n/2)\alpha]^{1/(1-2n)},~~~ n \not= 1/2, \eqno{(2.14b)}$$
$$(3)H=h= any~ value~ if~ \alpha=\alpha_c \equiv 2/\sqrt 3 , ~~~n=1/2  \eqno{(2.14c)}$$
\\
Using Eqs. (2.12) and (2.13) we can determinate the signs of $dH/dt$ and $dh /dt$ in the neighborhoods of the de Sitter space-time of $H=h=H_D$ and thus determine the stability of the de Sitter state. It is then found that {\it the isotropic cosmologies with $n <1/2$ display inflationary behavior but those with $n > 1/2$ shall exhibit the deflationary behavior as found in Ref.9. When $n =1/2$, then $H = h = 0$ and $H = h = \infty$ is the attractor state for $\alpha < \alpha_c$, and $\alpha > \alpha_c$, respectively. }  The investigation appears to display the division into $n >1/2$ , $n = 1/2$, and $n < 1/2$,also discussed by Barrow [19.20].

C. Cosmic time

The solutions expressed as the trajectories on the phase plane do not explicitly depend on the cosmic time. However, through a simple analysis we can determine whether the proper time in a solution is finite or infinite. 

   Equation (2.12) can lead to

$$  \int^H {d \bar H \over  [(\alpha /2)[\bar H(\bar H +2 \bar h)]^n - \bar H] \bar W              } = \int^t d\bar t.  \eqno{(2.15)}$$

We then see that the left-hand side in the above equation will become infinite only if $H$ is on a fixed point, and one can show that {\it the cosmic time of a point on a trajectory corresponding to a solution will be infinite if the trajectory has already met a fixed point. On the contrary, the trajectory starting with diverge $H$ will have finite proper time.}

\section{ Exact Solutions}

   We now begin to solve Eqs. (2.12) and (2.13). After dividing the former one by the latter one, we obtain

$${dH\over dt} = {\alpha [H(H+2h)]^n - 2H \over \alpha [H(H+2h)]^n - 2h}  .\eqno{(3.1)}$$
\\
The above equation can lead to

$${dL\over dt} = {3\alpha [HL]^n - 2L \over \alpha [HL]^n - 2h} , \eqno{(3.2a)}$$
$${dB\over dt} = {-2B \over \alpha [HL]^n - 2h} , \eqno{(3.2b)}$$
\\
where
$$L=H+2h, ~~~B=2(H-h).   \eqno{(3.3)}$$
Dividing Eq. (3.2a) by Eq. (3.2b) one gets

$${dL\over dB} = {L\over B}\left[1-{\alpha \over 2}3^{1-n} L^{2n-1}(1+{B\over L})^n \right]. \eqno{(3.4)}$$
\\
Using a new variable
$$A=L/B,  \eqno{(3.5)}$$
then Eq. (3.4) gives a simple form as

$$dA/(A^2+A)^n = -(\alpha /2) 3^{1-n} B^{2n-2}dB.  \eqno{(3.6)}$$
\\
The variables are thus separated and the equation becomes integratable. Through the integration by part, the exact solutions can be found when $2n$ is an integer.

   Since we will describe our solutions in the phase plane $H \times h$, we express $H$ and $h$ in the variables $r$ and $\theta$:

$$H= r cos\theta,                             \eqno{(3.7a)}$$
$$h= r sin \theta,                             \eqno{(3.7b)}$$
\\
in terms of which $A$ and $B$ become

$$A ={1+ 2 tan\theta \over 2(1 -  tan\theta)},   \eqno{(3.8)}$$
$$ B= 2r(cos \theta-sin\theta).                         \eqno{(3.9)}$$

Therefore, if Eq. (3.6) can be integrated exactly, the solutions that relate function $r$ to the variable $\theta$ will be found, and trajectories in phase plane can be plotted exactly, which in turn determine the evolutions of the cosmology. Various integration constants chosen in integrating Eq. (3.6) will produce the various trajectories which correspond to the solutions of the same model but with different initial states. The arrows in the trajectories, which tell us the directions of the evolution of the cosmology, are easily determined from Eqs. (2.12) and (2.13).

   As argued below, we also need to consider the solutions of negative energy density. A natural extension is the models with the viscosity function $\eta = \alpha (-\epsilon)^n$. Through the same procedure we can find the following relations:

$$dA/(-A^2-A)^n = - (\alpha/2)  3^{1-n} B^{2n-2}dB ~~~  (if ~~\epsilon \leq 0; H, W \geq 0),    \eqno{(3.10a)}$$
$$dA/(-A^2-A)^n =  (\alpha/2)  3^{1-n} B^{2n-2}dB ~~~  (if~~ \epsilon, H \leq 0;  W \geq 0),    \eqno{(3.10b)}$$
\\
The variables are separated and the exact solutions can be found when $n$ is an integer.

 As examples, we will give some explicit-solutions:

(1) $n=0$

$$ h-H_D=C(H-H_D), \eqno{(3.11)}$$
\\
where $C$ is an integration constant, and $H_D$ is defined in Eq.(2.14b). The solutions are thus all the straight lines that pass through the point of $H=h= H_D$  This means that the initial state of cosmology shall begin with $H$ and/or $h \rightarrow \infty$. Therefore, initial singularity will arise. However, we must be careful now.  The analyses in the above section show that the points on the line of $W = 0$ are the fixed points. Therefore some anisotropic solutions will begin on these points (in which $\epsilon$ is negative) at {\it infinite past}; during the evolution the cosmologies are isotropized and driven to the de Sitter state asymptotically. Using the relation of Eq. (3.11) we can from Eq. (2.12) find the function $H(t)$ which then explicitly shows this fact.

  Note that the energy density of these solutions is negative in the early stage; this will violate both the weak energy condition and the strong energy condition, and there is no singularity at any finite proper time. 

  Although it is difficult for the negative energy density to appear classically, it could be found in the quantized matter field. Also, as Hu [27] has discussed, the quantum dissipative process of the particle production could be formulated in terms of relativistic imperfect fluid.  Accordingly, it seem that there will be, more or less, some quantum senses in these solutions.  It is interesting to mention that the introducing of quantized matter field into the energy momentum tensor can sometimes lead to avoidance of the cosmological singularity, as found by Parker Fulling [28].

(2) $n =1/2$

$$r = C [ \pm(cos\theta - sin\theta) ]^{(\alpha_c /\alpha) -1} [\sqrt{cos \theta + 2 sin  \theta} + 3 \sqrt{cos \theta} ]^{2 \alpha_c / \alpha }  ~~~~~ (if ~~~ \epsilon \geq 0),
\eqno{(3.12a)}$$
\\
where $C$ is an integration constant and $\alpha_c$ is the constant defined in Eq. (2.14c). The solution shows that $H$ and/or $h$ can go to infinity only if $H = h$, i.e., $cos \theta = sin \theta $. Therefore, this model is with finite Hubble functions at the states of zero energy density ($H = 0$ or $H + 2h = 0$).

   The experiences from the analyses of the $n= 0$ model tell us that, as the points on the lines of zero energy density are not the fixed point, it is now also needed to discuss the regions of $\epsilon <0$.  (Note that we only discuss the expanding solutions, i.e., $W>0$.) We then must analyze the extended model in which the bulk viscosity is $\eta = \alpha (-\epsilon)^{1/2}$.  Equation (3.10) can give the explicit solutions:

$$r ={C\over sin\theta - cos\theta }~ exp\left[\pm {\alpha_c\over \alpha} sin^{-1} ({2+tan\theta\over 1- tan\theta})\right],    \eqno{(3.12b)}$$
\\
where C is an integration constant.

   After plotting the flows in the phase plane (see Fig. 1, in which, for clarity we adopt the non-uniform scale), we then find that anisotropic cosmologies shall always begin with finite negative energy density at the states with $H + h = 0$
in infinite past;  the Riemann scalar curvature and Hubble functions are finite in the initial phase. During the evolution, the energy density is increasing subsequently and the anisotropies of the universe are smoothed out. At the final stage as $t\rightarrow\infty$, depending on the value of $\alpha$ (and not on the value of $C$), there are three classes of states that may be approached asymptotically:
 1. Both the energy density and Hubble functions go to infinity if $\alpha >\alpha_c$.
2. Energy density is finite and space-time is attracted to a de Sitter universe (which is a function of the value of $C$)  if $\alpha =\alpha_c$. 
3. Both the energy density and Hubble functions decrease to approach zero and the model is driven to the isotropic Friedman universe $\alpha <\alpha_c$.

   The last two cases provide us with the solutions that are free cosmological singularity for all finite proper time.

(3) $n=1$

$$ r= [\alpha (cos\theta - sin\theta]^{-1}[C- ln(1+2tan\theta)],   \eqno{(3.13)}$$
\\
where $C$ is an integration constant. After plotting the flows in the phase plane (see Fig. 2, where, for clarity, we adopt the nonuniform scale), we then find that the anisotropic cosmologies shall start from the vacuum states and end in another
fixed point or infinite expansion state. It is then found that, except in the isotropic model ($h = H$) that has been investigated by Murphy [1], the cosmologies shall always begin with zero energy density at the initial phase of singularity.  During
the evolution, the energy density is increasing subsequently and the anisotropies of the universe are smoothed out.  At the final stage as $t\rightarrow\infty$, depending on the integration constant C (and not on the value of  $\alpha $), there are three classes of state that may be approached asymptoptically:

\vspace{3cm}
\special{jpg:bulk-1.jpg x=3cm y=3cm}

FIG. 1. The phase plane trajectories of the $n=1/2$: All solutions will begin with $W=0, \epsilon < o$ at $t\rightarrow -\infty$.

\vspace{3cm}
\special{jpg:bulk-2.jpg x=3cm y=3cm}

FIG. 2. The phase plane trajectories of the $n = 1$ model.  The solutions that evolve into positive Hubble functions at later stage will begin with $\epsilon = 0$ at finite past. 
\vspace{1cm}

1. Both the energy density and Hubble functions go to infinity if $H> h$, $C > ln 3$ or $h > H$, $ln 3 > C> 0$.

2. Energy density is finite and space-time is attracted to a de Sitter universe [which is determined by Eq. (2.14b) ] if $C=ln3$.

3. Both the energy density and Hubble functions decrease to approach zero and model is driven to the isotropic Friedmann universe if $H> h$, $In 3 > C > 0$ or $h > H$, $C > In 3$.

   The solutions that start at fixed point on the line of $W= 0$ are also described by Eq. (13) if one lets $\alpha \rightarrow -\alpha$. However, the trajectories that begin with $W= 0$ do not go into positive Hubble functions states at the final stage. 

(4) $n = 3/2$
$$r^2  = {1\over \alpha (cos\theta - sin\theta)^2 }\left[{4(2+tan\theta)\over \sqrt{1+2tan\theta}}-c\right]. \eqno{(3.14)}$$

This model can be used to describe the quantum production of infinitely thin Witten strings [23] on super-horizon scales in the very early universe [24,10].

(5)$n=2$
$$r^2  = {1\over \alpha (cos\theta - sin\theta)^2} \left[{3(2+tan\theta)(1-tan\theta)\over 1+2tan\theta}+{9\over2}(1+2tan\theta) -c\right]. \eqno{(3.15)}$$
\\
and so on.

    All the models of $n> 1$, as will be proved in the next section (for any $n$), possess the same characteristics such as the isotropization of the cosmology, beginning with zero energy density and with infinite Riemann scalar curvature, cre ating the matters in the course of evolution, and having three classes of final state.

\section{Analyses of Initial and Final States}

   We give in this section the analyses of the initial and final states. The results can clarify the characteristics of the cosmological evolutions for the models of any values of $n$. 

A. Initial states
The key equation to analyze the initial and final states of our model is Eq. (3.6). It can lead to

$$C_1 \int^A (\bar A^2+\bar A)^\sigma d\bar A + C_2(2A+1)(A^2+A)^\sigma + C_3 {(2A+1)\over (A^2+A)^{1-\sigma}} + C_4 {(2A+1)\over (A^2+A)^{n-1}} $$
$$+ C_5 {(2A+1)\over (A^2+A)^{n-2}} + ...= C_0 B^{2n-1}= C_0[2r(cos\theta-sin\theta)]^{2n-1},   \eqno{(4.1)}$$
\\
where $C_i$ are constant numbers which only depend on $n$, and $\sigma$ is chosen to satisfy $0 < \sigma  < 1$. It is important to notice that $C_3$ is nonzero only if $n \geq 1$, $C_4$ is nonzero only if $n \geq 2$,..., and so on.   If the universe is in the initial state $H + 2h \rightarrow 0$, then $A \rightarrow 0$ and $A^2 + A \rightarrow 0$, and we can easily prove that the first term (neglects the integration constant) and second term on the left-hand side of Eq. (4.1) shall approach zeros while the other terms become infinite. Therefore the value of $r$ in Eq. (4.1) is finite for the states that have vanish energy density, if and only if $0<n< 1$.

   However, it is now necessary to discuss the solutions in the regions of $\epsilon \leq 0$, as the $\epsilon = 0$ state is not the fixed point on the phase plane. From Eq. (3.10) one can show that $r$ is finite as $W \rightarrow 0$.  Hence we have proved that {\it the cosmologies of $0 < n < 1$ shall begin with finite negative energy density and zero total expand function at infinite past . They will then go into the positive energy density state}. Although {\it the models of $n > 1$ shall begin with zero energy density and diverge values of  $H$ and $h$ }, one can from Eq. (2.15) conclude that {\it the models with $n> 1$ shall begin at finite past}. 

   Furthermore, for the $n geq>l$ models we can from Eqs. (2.5) and (2.6) prove that near the initial phase the curvature singularity is the Kasner type, although the energy density is zero. This fact was first found in the letter of Belinskii and Khalatnikov [29], in which only the anisotropic n = 1 was analyzed.

{\bf B. Final states}
To analyze the final states we consider three cases separately:

Case 1: $n > 1$
The key equation can lead to
$$C_1 \int^A {d\bar A\over (\bar A^2+\bar A)^\sigma} + C_2{(2A+1)\over (A^2+A)^{n-1}} + C_3 {(2A+1)\over (A^2+A)^{n-2}} +...= C_0[2r(cos\theta-sin\theta)]^{2n-1}, \eqno{(4.2)}$$
\\
where $C_i$  are constant numbers that depend only on $n$, and $\sigma$ is chosen to satisfy $1 < \sigma < 2$. The value of $C_2$ is nonzero only if $n \geq 2$, $C_3$ is nonzero only if $n \geq 3$,..., and so on.   As universe approaches isotropic state, i.e., $H \rightarrow h$, then $A \rightarrow \infty $ and $B \rightarrow 0$, which in turn implies that the first term (neglects the integration constant) on the left-hand side ofEq. (4.2) becomes zero. Therefore, depending on the chosen integration constant, $r$ may be $?? \infty$. This situation is like that in the model of $n=l$. Therefore, depending on the initial state, the cosmology may be driven to infinite expansion state, de Sitter space-time, or isotropic Friedmann universe at the final stage.

Case 2: $ 1/2 < n < 1$

The key equation can lead to

$$C_1 \int^A {d\bar A\over (\bar A2+\bar A)^\sigma} + C_2{(2A+1)\over (A2+A)^n }= C_0[2r(cos\theta-sin\theta)]^{2n-1}, \eqno{(4.3)}$$
\\
where $C_i$, are constant numbers that depend only on $n$, and $\sigma = n + 1$, thus $1 <\sigma <2$. Using the arguments like that in Case 1. we can also show that the models with $1 > n > 1/2$  have three classes of final states. Hence we have proved that, {\it depending on the initial state, the cosmologies for the models of $n>1/2$  may be driven to infinite expansion state, de Sitter spacetime, or isotropic Friedmann universe at the final stage}. The model of $n =1/2$ was discussed in Sec. III.

Case3: $n <1/2$

The key equation can lead to

$$-{n\over 2(n-1)} \left[ \int_A {d\bar A\over (\bar A^2+\bar A)^{n+1}} + {(2A+1)\over n (A^2+A)^n}\right] = {-3^{1-n}\alpha \over 2(2n-1)}B^{2n-1}. \eqno{(4.4)}$$
\\
As universe approaches isotropic state, i.e., $H= h$ (it implies $A\rightarrow \infty $  and $B\rightarrow  0$), it is easy to prove that the first term in the bracket becomes zero (neglects the integration constant) and

$$r \rightarrow [3^n 2^{-(n+1/2)} \alpha]^{1/(1-2n)}, \eqno{(4.5)}$$
\\
no matter what the value of integration constant that will be chosen.  The state corresponding to Eq. (4.5) can be easily checked to be just the fixed point defined in Eq. (2.14b). Hence, we have proved that {\it solutions of the models of $n <1/2$ shall always be attracted to an isotropic de Sitter state at final stage}. 

  For the $n \leq 1$ models there are the solutions that start with $W= 0$ at infinite past and then are attracted to the original point on the phase plane. However, they never go into the state of positive Hubble function.

\section{ Models with Multiple Hubble Functions}

The methods described in the above sections can only be used to study the models with two Hubble functions. We will now give a simple algorithm that can be used to analyze the models with multiple Hubble functions. 

   Let us consider the $D + 1$-dimensional Binachi type I models. The Einstein's  field equation Eq. (2.2) can lead to

$${dH_i\over dt} + H_i W = (\epsilon -\bar p)/2,~~~ i=1,2,...D,  \eqno{(2.5)}$$
$$W^2 - \sum_i H_i^2 =  \epsilon,  \eqno{(5.2)}$$
\\
where $H_i$, are the Hubble functions and $W$ is the total expansion function. With the relation Eq. (2.9), and letting $\gamma = 2$ in there, we can from Eq. (5.1) find

$${dln(H_i-H_j)\over dt} = {dln(H_i-H_k)\over dt}. \eqno{(5.3)}$$
\\
This equation tells us that one can express all other $D-2$ Hubble functions in terms of only two Hubble functions. One can see that this is a very general property and may be used to analyze many other anisotropical cosmological models As examples, we will now describe the procedure of how to determine the three Hubble functions in the four-dimensional model. 

   Equation (5.3) can yield a relation
$$H_3 = (1-C)H + C h,                      \eqno{(5.4)}$$
in which $C$ is an integration constant, and for simplification, $H_1$, and $H_2$ are denoted as $H$ and $h$. Using Eq. (5.4) we can from Eq. (5.2) obtain

$$\epsilon = (1 - C)H^2 +2Hh + C h^2 = KL,         \eqno{(5.5)}$$
$$K= [(1-C)H + a h],                    \eqno{(5.6)}$$
$$ L= [H+b h],                             \eqno{(5.7)}$$
\\
where the constants $a$ and $b$ are the functions of $C$. Substituting the relation (5.5) into the viscosity function Eq. (2.11), then the Einstein's field equations (5.1) lead to

$${dK\over dt} ={\alpha\over 2}(1-C + a)(KL)^n W - KW,       \eqno{(5.8)}$$
$${dL\over dt} ={\alpha\over 2}(1+b)(KL)^n W - LW.       \eqno{(5.9)}$$
\\
Dividing Eq. (5.8) by Eq. (5.9), one gets

$${dK\over dL} ={\alpha(1-C + a)(KL)^n - 2K \over \alpha (1+b)(KL)^n - 2L}.       \eqno{(5.10)}$$
\\
After defining the variables

$$A = L/B,  \eqno{(5.11a)}$$
$$B =(1+b)K - (1-C + a)L, \eqno{(5.11b)}$$
\\
Eq. (5.10) gives a simple form

$${dA\over [A^2+(A/(1-C+a)]^n} =  {\alpha (1-C+a)\over 2 B^{2n-2}} dB.           \eqno{(5.12)}$$
\\
The variables are now separated and the equation is Integratable. (One can prove that $1 - C +a$ is nonzero.) Using the methods described in the above sections we can therefore analyze any dimensional Bianchi type I cosmological models with energy density dependent bulk viscosity. The results show the same characteristics as those in the models with two Hubble functions.

\section{CONCLUSIONS }

   We have analyzed in detail the anisotropic cosmological models with bulk viscosity ($\eta$) which is a power-law dependence upon energy density ($\epsilon$), i.e., $\eta = \alpha |\epsilon|^n$, when the universe is filled with stiff mattery $p=\epsilon$. We are interested in the cosmological solutions that will eventually go to the states of positive Hubble functions in the latter stage. Although the exact solutions could be obtained only when the $n$ is an integer, we are able to clarify the characteristics of evolution for the models of any $n$. 

   Let us give a summary. (1) There have been two kinds of solutions in the $n = 0$ model, which start either with diverge Hubble functions and infinite energy density at finite past or with finite Hubble functions and negative energy density at infinite past. However, both solutions are driven to a de Sitter space-time asymptotically. (2) All the solutions in the $0<n <1/2$ models will start with finite Hubble functions and negative energy density at infinite past and then are driven to a de Sitter space-time asymptotically. (3)  For the $1/2 \leq n < 1$ models, the initial state is with finite Hubble functions and negative energy density at infinite past; however, they can go to the infinite expansion state, de Sitter space-time, or Friedmann universe at final stage. (4) The $n \geq 1$ models will always begin with Kasner-type curvature singularity at final past in which the energy density is zero, however; and then they are driven to the above-mentioned three kinds of states asymptotically. All the solutions that begin with $\epsilon$ < 0 and then are attracted to a deSitter spacetime or Friedmann universe are free of cosmological singularity for any finite proper time.

    Historically, Murphy [8] presented the exact solution of the $n = 1$ model, and showed that the bulk viscosity can eliminate the big band singularity at any finite proper time. Belinskii and Khalatnikov [29] then analyzed the $n = 1$ Bianchi I model; they found that the cosmology is with the vanish energy density in the initial phase in which the Kasner singularity will arise. Now. the investigation in this paper shows that all the $n > 1$ Bianchi I models will share the same characteristics of initial singularity, and that the $n < 1$ Biancm models could give us some cosmological solutions that are free of singularity for all finite proper time. However, these singular-free solutions have negative energy density in the early epoch. 

   The models discussed in this paper are only for the stitt matter As the case of stiff matter is special because the shear and matter density behave in the same way in the absence of viscosity and vacuum and nonvacuum perfect fluid solutions are formally similar, the same models, while with other matter fields are certainly interesting, remain to be studied. 

   Finally, we want to mention that the prescription adopted in this paper can also be used to analyze the Bianchi type I cosmological models with energy density dependent shear viscosity [30]. The details will be presented elsewhere.

\newpage
\begin{enumerate}
\item  W.  Misner, Nature 214. 40 (1967); Astrophys. J. 151. 431 (1968).
\item  S. Weinberg, Astrophys. J. 168, 175 (1972).
\item   S. Weinbcrg, Gravitation and Cosmology (New York, Wiley, 1972).
\item  P. Langacker, Phvs.Rep.72. 185 (1981).
\item  L. Waga, R. C. Falcan, and R. Chanda,  Phys. Rev. D 33 1839 (1986).
\item  T. Pacher, J. A. Stein-Schabas, and M. S. Turner, Phys. Rev. D 36, 1603 (1987).
\item A. H. Guth. Phys. Rev. D 23, 347 (1981).
\item G. L. Murphy, Phys. Rev. D 8, 4231 (1973).
\item  N. O. Santo, R. S. Dias, and A. Banerjee, J. Math. Phys. 26, 876 (1985).
\item J. D. Barrow, Nucl. Phys. B. 310, 743 (1988).
\item V.A. Belinski and I. M. Khalatnikov, Sov. Phys. JETP 42, 205 (1976)
\item A. Banerjee and N. O. Santos,  J. Math. Phys. 24, 2689 (1983).
\item A. Banerjee and N. O. Santos, Gen. Relativ. Gravi. 16. 217 (1984); 18,
461 (1986).
\item A. Banerjee, S. B. Duttachoudhury, and A. K. Sanyal, J. Math. Phys. 26,
3010(1985).
\item G. W. Gibbons and S. W. Hawking, Phys. Rev. D 15, 2738 (1977).
\item S. W. Hawking and I. G. Moss. Phys. Lett B. 110, 35 (1982).
\item  R. M. Wald, Phys. Rev. D 28, 2118 (1983); L. G- Jensen and A. Stein-
Schabes, Phys. Rev. D 35, 1146 (1987).
\item S. W. Hawking and G. F. R. Ellis, The Large Scale Structure Of Spacetime
(Cambridge U.P., Cambridge, 1973).
\item J. D. Barrow, Phys. Lett. B. 180, 335 (1986).
\item J. D. Barrow, in The Early Universe, edited by W. Unroh and G. W. Se-
menoff (Reidel, Dordrecht, 1988), pp. 191-194. 
\item W. -H. Huang, Phys. Lett. A. 129, 429 (1988).
\item J. D. Barrow, Phys. Lett. B. 183, 285 (1987).
\item  E. Witten, Phys. Lett. B. 153, 243 (1985).
\item N. Turok, Phvs. Rev. Lett. 60, 548 (1988).
\item Y. B. Zeidorich, Sov. Phys. JETP 14, 1143 (1962).
\item J. D. Barrow, Nature 272, 211 (1978).
\item B. L. Hu, Phys. Lett. A 90, 375 (1982).
\item  L. Parker and S. A. Fulling, Phys. Rev. D 7, 2357 (1973).
\item  A. Belinskii and I. M. Khalatnikov, Sov. Phys. JETP Lett. 21, 99
(1975).
\item  W. -H. Huang, "Effects of the shear viscosity on the character of  cosmological evolution," J. Math Phys. 31, 659 (1990). [gr-qc/0308060]
\end{enumerate}

\end{document}